\documentclass[sigconf,screen]{acmart}
\AtBeginDocument{%
  \providecommand\BibTeX{{%
    \normalfont B\kern-0.5em{\scshape i\kern-0.25em b}\kern-0.8em\TeX}}}

\usepackage{multicol}
\usepackage{color, colortbl}
\definecolor{Gray}{gray}{0.9}

\usepackage{draftwatermark} 

\begin{document}
\title{Human Aspect of Threat Analysis: A Replication}

\author{Katja Tuma}
\email{k.tuma@vu.nl}
\affiliation{%
  \institution{Vrije Universiteit Amsterdam}
  \country{The Netherlands}
}
\author{Winnie Mbaka}
\email{w.mbaka@vu.nl}
\affiliation{%
  \institution{Vrije Universiteit Amsterdam}
  \country{The Netherlands}
}

\renewcommand{\shortauthors}{K Tuma and W Mbaka}

\begin{abstract}
\textbf{Background:} 
Organizations are experiencing an increasing demand for security-by-design activities (e.g., STRIDE analyses) which require a high manual effort.
This situation is worsened by the current \textit{lack of diverse (and sufficient)} security workforce and inconclusive results from past studies.
To date, the deciding human factors (e.g., diversity dimensions) that play a role in threat analysis have not been sufficiently explored.\\
\textbf{Objective:} To address this issue, we plan to conduct a series of exploratory controlled experiments. The main objective is to empirically measure the human-aspects that play a role in threat analysis alongside the more well-known measures of analysis performance. \\
\textbf{Method:} We design the experiments as a differentiated replication of past experiments with STRIDE. The replication design is aimed at capturing some similar measures (e.g., of outcome quality) and additional measures (e.g., diversity dimensions). We plan to conduct the experiments in an academic setting.\\
\textbf{Limitations:} Obtaining a balanced population (e.g., wrt gender) in advanced computer science courses is not realistic. The experiments we plan to conduct with MSc level students will certainly suffer this limitation. 
\end{abstract}


\maketitle
\SetWatermarkText{Pre-print} 
\SetWatermarkScale{0.5} 

\section{Introduction}
\label{sec:intro}
Security-by-design techniques~\cite{DSS+2009,STM2017} have been used to prevent costly security fixes to software in later stages of the development life-cycle by analyzing security already during the  design phase.
Practitioners use threat analysis~\cite{tuma2018threat} to look for potential security threats in their product's software architecture.
For instance, STRIDE~\cite{shostack2014threat} is a popular technique developed by Microsoft.

There is an increasing need to perform such architectural security analyses (e.g., latest BSIMM study reports an increased investment by more than 65\%~\cite{BSIMM12:online}) as the threat landscape evolves.
However, threat analysis requires a high manual effort~\cite{scandariato2015descriptive}, demands the involvement of security and domain experts~\cite{cruzes2018challenges}, and has been proven time and again difficult to fully automate~\cite{tuma2020automating}.

Threat analysis practices are set back by a globally recorded shortage of the security workforce~\cite{cybersec2019,blavzivc2021cybersecurity}.
In addition, the current security workforce is not diverse (e.g., with respect to gender) which may be viewed as an opportunity for a change. 

Risk decisions (which are core to threat analysis) are made in face of uncertainty~\cite{bier2020role}, thus there is space for subjective (and possibly biased) judgement~\cite{jaspersen2015probability,brito2020predicting}.
Empirical evidence of threat analysis performance indicators is a crucial piece of the puzzle to improve the situation.
But, past empirical studies were either inconclusive about some performance indicators~\cite{tuma2018two} or have focused on measuring performance indicators irrespective of the human factors~\cite{scandariato2015descriptive,tuma2021finding,van2021descriptive}. 
Yet measuring such human factors is pivotal to understanding how to close the security workforce gap in the future.

To address these issues, we plan to conduct a series of exploratory controlled experiments with the aim of empirically measuring the human-aspects that play a role in threat analysis. 
In particular, we design a differentiated replication~\cite{lindsay1993design}, where we capture some similar measures used in previous experiments~\cite{tuma2018two} but also different measures (e.g., participant gender, nationality, type of outcomes). 


\section{Related Work}
\label{sec:related-work}
We positioned our contributions with respect to existing literature on empirical studies of STRIDE and related replication studies.\\
\textbf{Empirical studies of threat analysis.} 
Several works have investigated STRIDE empirically. 
Scandariato et al.~\cite{scandariato2015descriptive} performed a descriptive analysis to quantify the cost and effectiveness of STRIDE by measuring the productivity, precision, and recall in an academic setting. 
Their study reports students having higher rates in precision and recall with lower productivity rates. 


Two studies~\cite{cruzes2018challenges,bernsmed2019threat} conducted case studies investigating the challenges of STRIDE.
Bernsmed et al.~\cite{cruzes2018challenges} performed an exploratory case study with the goal of investigating the challenges facing adoption of threat modeling using the Microsoft approach with STRIDE. The study was done in a company comprising five agile development projects. Their analysis elicited 21 challenges to threat modelling which were then mapped to existing literature and concluded that proper understanding of threat elicitation is required in order to actualise the functions of STRIDE especially in agile development.


Stevens et al. \cite{stevens2018CoG} conducted a qualitative case study to  evaluate the impact of introducing threat modeling to an organization that had previously not used it by applying the Center of Gravity (CoG) framework.
The CoG is a risk-first threat analysis technique that has not been extensively used to analyze software security. The authors goal was to measure effectiveness and efficiency of CoG, using surveys and classroom sessions that involved 25 practitioners. The authors reported a very high accuracy from the participant's results, similar to other studies that have been conducted with experts.
Different attack and risk-centric threat analysis techniques (such as Attack trees, CORAS, MUCs, to name a few) have also been investigated with empirical rigour. We point the interested reader to a systematic review for more details~\cite{tuma2018threat}.\\
\textbf{Replications.} We frame our plan as a series of experimental replications.
Generally, the goal of replication studies is to validate the experimental procedures of the original study using a different participation pool.
This studies are aimed at generating new data~\cite{santos2019procedure} (as opposed to re-analyzing the same data in reproduction studies). 
We briefly mention some related replication studies.

In their original study, Labunets et al~\cite{labunets2013experimental}, compared two risk assessment methods, a visual and a textual method and reported that the visual method was more effective for identifying threats than the textual one. The same study was replicated in ~\cite{labunets2017equivalence}, applying similar procedures. In contrast to the original study, the replication reported that the two methods being investigated were (statistically) equivalent with regards to the quality of identified threats and security controls.



Several studies have empirically compared~\cite{opdahl2009experimental,karpati2011experimental,karpati2012comparing} and conducted replications~\cite{Jung2013223,Riaz20172127,Rueda2020} requirement engineering techniques (e.g., requirements elicitation). 
For brevity, we direct the interested reader a comprehensive review by Ambreen et al.~\cite{ambreen2018empirical}.


\section{Research questions}
\label{sec:rqs}

Due to the academic setting we limit this study on observing gender, background, and nationality diversity dimensions (and exclude seniority).
The main goal of this study is to measure the existence (or absence) of diversity effects on the actual and perceived analysis outcomes.
Accordingly, we developed two research questions and hypotheses about each measure.

\smallbreak
\noindent
\textbf{RQ1.} \textit{What is the effect of gender, background, and nationality on the \textbf{actual} threat analysis outcomes?}
\smallbreak
To investigate RQ1, we pose hypotheses about the \textbf{equivalence of the sample means for the analysis outcomes}.

\noindent
$H1_1: Comp. Sci_F = Comp. Sci_M$
\smallbreak

\noindent
Regarding gender, we expect that the outcomes reported by women are equivalent to the outcomes reported by men. 
Studies of risk perception suggest that women perceive certain risks differently compared to men.
Though we do not foresee strong differences, we might find some effects when it comes to risk priority.

\smallbreak
\noindent
$H1_2: Comp. Sci_1 = Comp. Sci_2$
\smallbreak

\noindent
Regarding education, we expect that the students of various specialization tracks report equivalent outcomes for the same system under analysis.

\smallbreak
\noindent
$H1_3: Comp. Sci_{Na} = Comp. Sci_{Nb}$
\smallbreak

\noindent
We expect that the students of various race and nationality report statistically equivalent outcomes. 

\smallbreak
\noindent
\textbf{RQ2.} \textit{What is the effect of gender, background, and nationality on the \textbf{perceived} threat analysis outcomes?}
\smallbreak
To investigate RQ2, we pose hypotheses about the \textbf{equivalence of the sample means for the perceived analysis outcomes}.

\noindent
$H2_1: Perc(Comp. Sci_F) < Perc(Comp. Sci_M)$
\smallbreak

\noindent
Due to low confidence levels of female computer science students, we expect that the perceived quality of outcomes reported by women is lesser compared to the perceived quality of outcomes reported by men (regardless of the actual outcomes by both groups). 

\smallbreak
\noindent
$H2_2: Perc(Comp. Sci_1) = Perc(Comp. Sci_2)$
\smallbreak

\noindent
Regarding education, we expect that overall the students of various specialization tracks do not differ in their perceived quality of the outcomes they produced.
We may find higher confidence levels of perceived quality for students that are following a security specialization track.

\smallbreak
\noindent
$H2_3: Perc(Comp. Sci_{Na}) = Perc(Comp. Sci_{Nb})$
\smallbreak

\noindent
We expect that the students of various nationality do not differ in their perceived quality of the outcomes they produced.



\section{Replication protocol}
\label{sec:intro}

\subsection{Variables}
\label{subsec:vars}
Table~\ref{tab:vars} shows the variables of the study.

\subsubsection{Independent} 
\textbf{Gender} is an individual's identity based on their sex, which is typically, man and woman, but can also be non-binary.
Several studies ~\cite{rodriguez2021perceived} ~\cite{imtiaz2019investigating} ~\cite{wang2019implicit} found evidence of bias against women in some software engineering communities, and sometimes negative perceptions about women working in teams.
Thus, this is an interesting dimension to further investigate in the context of security.\\
\textbf{Education} is an individuals' level of achievement within a specific area of specialization (e.g., computer security vs AI) in academic studies. Typically, risk-based decisions in organizations have been made by persons in managerial positions who are assumed to have a better understanding of the product. However, technical skills of security experts or engineers should be taken into consideration during critical decision-making. Therefore, it is interesting to investigate the effect of education by including participants from different academic backgrounds (e.g., in communication sciences).
Although in ~\cite{jardine2020case} study, education was found to be a non-significant variable, it is not clear whether this dimension has an impact in performing a RA task. \\
%
\textbf{Nationality} is generally used to refer to someone's country of origin, however, it may be coupled with other identifying aspects, such as the culture and language. On the other hand, race is a social construct that is associated with an individuals physical appearance, such as their skin colour. 
Determining the effect of nationality bias in security practices is to date an open question.
In a previous study, Thomas et al.~\cite{thomas2018speaking} conducted semi-structured interviews with 11 Black women in computing and report that Black women experienced a number of challenges, such as discrimination, expectations from others that are too high or too low, isolation, sexism, and racism. However, it is not clear whether some of these challenges were as a result of their gender, race or nationality).
But, few studies have focused on nationality diversity in the software engineering discipline~\cite{rodriguez2021perceived}.

\subsubsection{Dependent.} 
Since the quality of analysis lacks a formalised definition (e.g., often natural language is used to describe attack scenarios and informal notations are used for modeling~\cite{tuma2018threat}), we will use measures that can be easily reproduced.
Namely, we can observe how diversity dimensions effect the \textit{type of analysis outcomes}.
Table \ref{tab:vars} (dependant variables) shows various outcomes types that we observe. \\
\textbf{Threats.} We use the STRIDE threat categories to distinguish different type of threats. Tuma et al.~\cite{tuma2018two,tuma2021finding} noticed that expert analysis tend to be more balanced in terms of their review of different threat categories, while non-experts tend to report a high number of tampering, denial of service and information disclosure threats.
We are interested to observe whether other diversity dimensions have an effect on category distribution of identified threats.\\
\textbf{Assumptions.} Assumptions are statements about the domain that may or may not be true. Assumptions are often implicit and dynamic in nature (i.e., they can be invalidated and modified as the project evolves).
Van Landuyt and Joosen~\cite{van2021descriptive} find that the majority of assumptions (created by students during STRIDE) were used to either justify an existence of threats or are used to eliminate threats.
In~\cite{van2021descriptive} a substantial subset (78\%) of the assumptions was in direct reference to security-related concepts (i.e., security assumptions), however also domain assumptions (statements about component functionalities) were made.
Thus, we are interested to investigate the effect of diversity dimensions on the type of assumptions.\\
\textbf{Attack surface.} defining an attacker profile and the \textit{attack surface} are essential in determining the feasibility of an attack scenario. To this end, security analysis are expected to make these distinction prior to performing a threat analysis. \\
%
\textbf{Risk priority.} We refer to risk as a product of threat probability and impact. How individuals assess risk priorities may be related to their risk perception which is already well understood~\cite{gustafsod1998gender}. Since the number of identified threats explodes in realistic projects, practitioners must choose which threats are most urgent to mitigate. 
Thus they prioritize them based on estimations of risk.\\
\textbf{Mitigations.} There are different approaches that can be applied while mitigating security risks.  Preventative (e.g., implementation of two-factor authentication), detective/reactive (e.g., using intrusion detection and access revocation techniques) and corrective (e.g., maintaining audit trails or restoring from a secure state). Security analysts can implement multiple strategies depending on domain-related factors, such as the cost associated with a specific mitigation strategy. Some diversity dimension (e.g., gender)
may underestimate the ease with which a mitigation is actually implemented, as observed in~\cite{wright2002empirical}. 
Thus, it is interesting to observe how other diversity dimensions, including gender effect the type of mitigations that are identified during threat analysis.

\begin{table*}[]
    \centering
    \begin{tabular}{p{0.3\textwidth} p{0.4\textwidth} p{0.05\textwidth} p{0.15\textwidth}}
        \toprule
         \textbf{Name} & \textbf{Description} & \textbf{Scale} & \textbf{Operationalization} \\
         \midrule
         \multicolumn{4}{l}{\textit{Independent variables (design)}} \\
         \midrule
         Gender & obtained from the gender of participants & nominal &  multiple choice \\
         Background & the program specialization and extra curriculum activities & nominal &  multiple choice \\
         Nationality & obtained from the nationality of participants & nominal &  multiple choice \\
         \midrule
         \multicolumn{4}{l}{\textit{Dependent variables}} \\
         \midrule
         \multicolumn{4}{l}{\textit{**Different measures compared to existing literature**}} \\
         \rowcolor{Gray}
         Type of identified threats & distribution of categories of threats (spoofing, tampering, information disclosure, denial of service, elevation of privilege) that have been identified by the participants & nominal & see Section~\ref{subsec:tasks} \\
         \rowcolor{Gray}
         Type of assumptions & distribution type of assumptions (domain, security) that have been reported by the participants & nominal & see Section~\ref{subsec:tasks} \\
         \rowcolor{Gray}
         Type of attacks surface & distribution attack surfaces (physical, close-proximity, remote) of the identified threats & nominal & see Section~\ref{subsec:tasks} \\
         \rowcolor{Gray}
         Risk priorities & distribution of risk priorities (high, medium, low) assigned to identified threats & nominal & see Section~\ref{subsec:tasks} \\
         \rowcolor{Gray}
         Type of mitigations & distribution of type of identified mitigations (preventative, detective/reactive, corrective) & nominal & see Section~\ref{subsec:tasks} \\
         \midrule
         \multicolumn{4}{l}{\textit{Treated/Measured variables}} \\
         \midrule
         Time spent on task & time (in hours) to complete the task using the prescribed technique & ordinal & automatically measured by the submission tool \\
         Perceived precision (PP) & self-reported ratio between the number of correctly identified threats and all \textit{threats identified} & ordinal &  5-point Likert scale \\
         Perceived recall (PR) & self-reported ratio between the number of correctly identified threats and all \textit{existing} threats identified  & ordinal &  5-point Likert scale \\
         Perceived usefulness (PU) & self-reported usefulness of the prescribed technique & ordinal &  5-point Likert scale \\
         Experience with security and modeling & self-reported experience in number of years or previously completed courses & ordinal &  5-point Likert scale \\
         Experience with STRIDE & self-reported experience in number of years or previously completed courses  & ordinal &  5-point Likert scale \\
         Experience with domain of application & self-reported experience in number of years or previously completed courses  & ordinal &  5-point Likert scale \\
         \rowcolor{Gray}
         \multicolumn{4}{l}{\textit{**Different measures compared to existing literature**}} \\
         \rowcolor{Gray}
         Perceived cognitive load & the reported cognitive load (complexity) of the task using the prescribed technique & ordinal & 5-point Likert scale \\
         \rowcolor{Gray}
         Perceived quality and efficacy & the reported quality and self efficacy of work conducted & ordinal &  5-point Likert scale \\
        \bottomrule
    \end{tabular}
    \caption{Variables of the differentiated replication experiments}
    \label{tab:vars}
\end{table*}

\subsection{Material}
\label{subsec:tasks}
\textbf{Training.} In the first part of the training the participants will be introduced to some key security topics (such as CIAA triad, security threats, attack surface and vulnerabilities, security controls and risk mitigations). The second part of the training will prepare the students to actually perform a threat analysis using one of the technique variants. The third part of the training will introduce the participants to the case study which will be the object of their analysis.\\ 
\textbf{Case study documentation.} We will use the same case study as in the original study. The home monitoring system (HomeSys) is an automated surveillance system designed for residential places. Its main objective is to enable the home-owner to remotely monitor their property. A detailed documentation of the case (requirements, architectural design, etc) will be made available to the participants. \\
\textbf{Ground truth analysis.} We will use one 'golden standard' data flow diagram and its' corresponding ground truth STRIDE analysis of the HomeSys case study from~\cite{tuma2018two}.
Since we do not aim to measure the quality of the diagrams created, and the DFD building is less time consuming compared to threat identification, we will provide a model to the participants.
This will significantly simply the comparison of the identified security threats.
Similarly, we will provide the ground truth analysis to the participants that will be prioritizing threats and identifying security mitigations.\\
\subsection{Task}
The participants will be asked to individually fill-in a survey.
The survey consists of three parts.
First, a few questions about the students gender, background, nationality.
Half of the participants will be asked to perform a STRIDE analysis (i.e., identify security threats).
In contrast to previous studies, our participants will \textit{be given the same graphical model} of HomeSys to analyze and they will analyze it using the same STRIDE technique.
The other half of the participants will be asked to prioritize a list of security threats and identify security mitigations to high-priority threats.
In contrast to previous studies, our participants will be given the graphical model \textit{and the list of security threats}.
To guide the threat identification the participants will use the documentation of STRIDE.
Similar to the past studies, we will hand out a threat template csv to standardize the format of the outcomes reported.
The participants will submit the files using the same survey.
Finally, they will be asked a few questions regarding their perception of the task.
Time taken to complete the task was captured using an online survey tool. 

\subsection{Participants}
\label{subsec:participants}
Our population is computer science students, with some differences in the elective courses and program choices (e.g., we plan to include students from various master programs, such as IA, computer Security, and Software Engineering).
All participants are students enrolled in a course taught by the experimenters. 
At the beginning of the course we plan to hand out an entry survey to measure participants' background and areas of expertise relevant to the study. 
We expect most to be new to secure design techniques (e.g STRIDE, threat modeling, Data Flow Diagrams, misuse cases, attack trees etc). 
In addition, we expect the participants are unfamiliar with architectural modeling techniques (e.g sequence, component and deployment diagrams).

\subsection{Execution plan}
\label{subsec:execution}
\textbf{Work division.} The participants will be randomly divided into two groups (A and B).
Group A will be tasked with analysing a provided data flow diagram of the HomeSys case study using STRIDE.
Group B will be tasked with prioritizing a provided list of security threats and identifying security mitigations for high-priority threats.
These treatment groups are formed only to divide the work to avoid overloading the participants performing an overly complex task individually.\\
\textbf{Training.} The participants will undergo an obligatory training lectures (about 3 hours) covering the topics mentioned above. \\
\textbf{Hand-outs.} After the training, participants will be given digital copies of all the support material (inc. lecture slides, case documentation, technique documentation, etc). \\
\textbf{Physical labs.} The experiment will be conducted during a four hour physical lab. The participants will be separated into different classrooms depending on their treatment group (to avoid spillover effects).
Each classroom will be supervised by either a teaching assistant or the experimenters. 
Only questions about the experiment protocol will be answered.\\
\textbf{Reports.} The data will be collected through an online survey tool.

\subsection{Analysis plan}
\label{subsec:analysis}
\textbf{Data cleaning.} We will perform a preliminary check of the collected data. This will include removing submissions for which we did not get explicit consent by the participant. Second, we will remove clearly insincere submission attempts (if any).\\
\textbf{TOST analysis of equivalence.} We will use both difference and equivalence statistical tests. 
It is likely that we do not obtain normally distributed samples, thus we plan to use a non-parametric, Mann-Whitney test. TOST was initially proposed by \cite{schuirmann1981hypothesis} and is widely used in pharmacological and food sciences to check whether two treatments are equivalent within a specified range $\delta$~\cite{food2001guidance,meyners2012equivalence}. 
Wherever possible (e.g., for Likert-scale questions) we will define the delta empirically.
For instance by pooled variance $\sigma_p$ across several samples reported in the literature on security risk analysis (e.g., in a four year interval) on variables ranging over a 5-item Likert scale for demographic statistics as to account for natural variability of the data. \\
\textbf{Validity threats.} There is typically around 20\% (or less) female students enrolled in computer science programs.
We are aware of the validity threats caused by an unbalanced population sample, which is omnipresent in all gender diversity studies in STEM disciplines~\cite{rodriguez2021perceived}.
To partially mitigate this threat, we will rally female computer scientist students towards participation through local steminist groups and similar community organized channels.
 
Since we do not include practitioners in this study, we can not observe the full complexity of the diversity effects (e.g., including seniority) that are actually present in organizations where threat analysis is routinely performed.
Still, studies have shown~\cite{runeson2003using, host2000using, salman2015students} that the differences between the performance of professionals and graduate students are often limited.

We considered the threat of overloading the participants with a complex task. We mitigate this threat by splitting the participants into two groups, so individual participants get to either only focus on finding threats or focus on mitigating risks.


\section{Acknowledgments}

\bibliographystyle{ACM-Reference-Format}
\bibliography{literature}

\end{document}